\documentclass[twocolumn,showpacs,preprintnumbers]{revtex4}
\usepackage{amssymb}
\usepackage{amsmath}
\usepackage{graphicx}
\usepackage{dcolumn}
\usepackage{bm}

\begin{document}

\title{Genuine tripartite entanglement monotone of $(2\otimes 2\otimes n)-$%
dimensional systems}
\author{Chang-shui Yu}
\author{L. Zhou}
\author{He-shan Song}
\email{hssong@dlut.edu.cn}
\affiliation{School of Physics and Optoelectronic Technology, Dalian University of
Technology, Dalian 116024, P. R. China}
\date{\today }

\begin{abstract}
A genuine tripartite entanglement monotone is presented for $(2\otimes
2\otimes n)$-dimensional tripartite pure states by introducing a new
entanglement measure for bipartite pure states. As an application, we
consider the genuine tripartite entanglement of the ground state of the
exactly solvable isotropic spin-$\frac{1}{2}$ chain with three-spin
interaction. It is shown that the singular behavior of the genuine
tripartite entanglement exactly signals a quantum phase transition.
\end{abstract}

\pacs{03.65.Ud, 03.67.Mn}
\maketitle

\section{Introduction}

The existence of quantum entanglement is a joint consequence of the
superposition principle and the tensor product structure of the quantum
mechanical state space of composite quantum systems. Entanglement is a
remarkable feature that distinguishes the quantum from the classical world.
One of the main tasks of quantum entanglement theory is to quantitively
characterize the extent to which composite quantum systems are entangled by
constructing a mathematical function------an entanglement measure that
should be an entanglement monotone, or in other words, not increase on
averaging under local operations and classical communications (LOCC). Even
though many efforts have been applied to a variety of quantum systems [1-3],
only bipartite pure-state and low-dimensional systems are well understood.
The quantification of entanglement for high-dimensional systems and
multipartite quantum systems remains an open question.

An important step in studying multipartite entanglement was taken by
Coffman, Kundu and Wootters [4]. They showed that a quantum state has only a
limited shareability for quantum entanglement, when they introduced the
so-called residual entanglement for tripartite systems of qubits based on
the remarkable concurrence [1] to measure an essential three-qubit
(three-way) entanglement which must be shared by all the three qubits. A
representative example with such a property is the
Greenberger-Horne-Zeilinger (GHZ) state which has maximal residual
entanglement. Once a qubit is traced out, the remaining two qubits are
separable. Later, it was shown that this essential three-qubit entanglement
is one (GHZ-type) of the two inequivalent classes (GHZ-type and W-type) of
tripartite entanglement of qubits [5]. In this paper, we call it genuine
tripartite entanglement. For a general multipartite quantum state,
multipartite quantum entanglement can be classified into much more
inequivalent entanglement classes [6-9], which is also one principal reason
why the entanglement in multipartite systems is much more complicated.
Therefore, in general a single quantity can not effectively and thoroughly
measure multipartite entanglement. However, sometimes a single quantity is
quite convenient and straightforward if one is going to study the
separability property of a given quantum system [10-12], or collect the
contributions of some entanglements of different classes as a whole [13-16],
or more naturally, measure entanglement of a given class [6-9].

In this paper, we introduce a single quantity to characterize the genuine
tripartite entanglement of tripartite $(2\otimes 2\otimes n)-$dimensional
quantum systems based on a new bipartite entanglement measure. The distinct
advantage is that the quantity is not only an entanglement monotone, but
also explicitly quantifies the GHZ-type inseparability of a tripartite
high-dimensional pure state. As an application, we consider the genuine
tripartite entanglement of the ground state of the exactly solvable
isotropic spin-$\frac{1}{2}$ chain with three-spin interaction [17,18]. It
is shown that the singularity of the genuine tripartite entanglement exactly
signals a quantum phase transition. The paper is organized as follows.
First, we introduce a new bipartite entanglement monotone; and then we give
the genuine tripartite entanglement monotone and consider the entanglement
of the ground state of the exactly solvable isotropic spin-$\frac{1}{2}$
chain with three-spin interaction; the conclusions are drawn at the end.

\section{A new entanglement monotone for bipartite pure states}

\bigskip As we know, a bipartite quantum pure state $\left\vert \psi
\right\rangle _{AB}$ defined in $\left( n_{1}\otimes n_{2}\right) $
dimension is, in general, considered as a vector, i.e. $\left\vert \psi
\right\rangle _{AB}=[a_{00},a_{01},\cdot \cdot \cdot
,a_{0n_{2}},a_{10},a_{11},\cdot \cdot \cdot ,a_{n_{1}n_{2}}]^{T}$ with
superscript $T$ denoting transpose operation. But in a different notation, $%
\left\vert \psi \right\rangle _{AB}$ can also be written in matrix form as
\begin{equation}
\psi _{AB}=\left(
\begin{array}{cccc}
a_{00} & a_{01} & \cdots & a_{0n_{2}} \\
a_{10} & a_{11} & \cdots & a_{1n_{2}} \\
\vdots & \vdots & \ddots & \vdots \\
a_{n_{1}0} & a_{n_{1}1} & \cdots & a_{n_{1}n_{2}}%
\end{array}%
\right) ,
\end{equation}%
by which one can easily find the reduced density matrix
\begin{equation}
\rho _{A}=\psi _{AB}\psi _{AB}^{\dag }.
\end{equation}%
Let $\sigma _{i}$, $i=1,2,\cdots ,\mathcal{D}$, with $\mathcal{D}=\min
\{n_{1},n_{2}\},$be the singular values of $\psi _{AB}$ or the square roots
of eigenvalues of $\rho _{A}$ in decreasing order. Define
\begin{equation}
\mathcal{S}\left( \rho _{A}\right) =\left( \sum\limits_{i=1}^{\mathcal{D}}%
\sqrt{\sigma _{i}}\right) ^{2},
\end{equation}%
then $E(\left\vert \psi \right\rangle _{AB})\ $is defined by
\begin{equation}
E(\left\vert \psi \right\rangle _{AB})=\mathcal{N}\left[ \mathcal{S}\left(
\rho _{A}\right) -1\right]
\end{equation}%
with $\mathcal{N}$ an adjustable constant by which one can select different
reference frames. For example, one can set $\mathcal{N=}\frac{2\sqrt{2}+1}{7}
$ such that $E=1$ for Bell states and $\mathcal{N=}\frac{\mathcal{D}\sqrt{%
\mathcal{D}}+1}{\mathcal{D}^{3}-1}$ such that $E=1$ for $\mathcal{D}$%
-dimensional maximal entangled states.

\textbf{Theorem 1}. $E(\left\vert \psi \right\rangle _{AB})$\textit{\ is an
entanglement measure of }$\left\vert \psi \right\rangle _{AB}$\textit{.}

\textbf{Proof.} It is obvious that $\mathcal{S}\left( \rho _{A}\right) =1$
(the rank of $\psi _{AB}$ is one) if $\left\vert \psi \right\rangle _{AB}$
is separable, i.e., $E(\left\vert \psi \right\rangle _{AB})=0$. Conversely,
if $E(\left\vert \psi \right\rangle _{AB})=0$, $\mathcal{S}\left( \rho
_{A}\right) =1$ which implies that the rank of $\psi _{AB}$ is one, i.e., $%
\left\vert \psi \right\rangle _{AB}$ is separable. These results show that $%
E(\left\vert \psi \right\rangle _{AB})=0$ is a sufficient and necessary
condition for separability. In fact, it is equivalent to the separability in
terms of the Schmidt decomposition.

Now, we show that $E(\left\vert \psi \right\rangle _{AB})$ is an
entanglement monotone, i.e., $E(\left\vert \psi \right\rangle _{AB})$ does
not increase under LOCC operations. At first, it is easily found that $%
E(\left\vert \psi \right\rangle _{AB})$ does not change under local unitary
transformations because the singular values of $\psi _{AB}$ are invariant
under such transformations. Next, without loss of generality, we suppose
that the local operations are only performed on the subsystem $A$.
Furthermore, for simplicity, we assume a local unitary transformation $Z$ is
performed on subsystem $A$ beforehand. This is valid because local unitary
transformations do not change the entanglement. Analogously to Ref. [5], let
$A_{1}$ and $A_{2}$ be two Positive-Operator-Value-Measurement (POVM)
elements such that $A_{1}^{\dag }A_{1}+A_{2}^{\dag }A_{2}=\mathbf{1}_{A}$,
with $\mathbf{1}_{A}$ denoting the identity of subsystem $A$ and $%
A_{i}=U_{i}D_{i}V$, where $U_{i}$ and $V$ are unitary matrices and $D_{i}$
are diagonal matrices with entries $(a_{1},a_{2},\cdots )$ and $[\sqrt{%
1-a_{1}^{2}},\sqrt{1-a_{2}^{2}},\cdots ]$, respectively. For some initial
state $\left\vert \psi \right\rangle _{AB}$, let $\left\vert \theta
_{i}\right\rangle =(A_{i}Z\otimes \mathbf{1}_{B})\left\vert \psi
\right\rangle _{AB}$ be the unnormalized states obtained after the POVM
operations. The corresponding normalized states can be given by $\left\vert
\theta _{i}^{\prime }\right\rangle =\left\vert \theta _{i}\right\rangle /%
\sqrt{p_{i}}$, where $p_{i}=\left\langle \theta _{i}\right\vert \left.
\theta _{i}\right\rangle $. Then the average entanglement after operations
can be given by

\begin{equation}
\left\langle E(\left\vert \psi \right\rangle _{AB})\right\rangle
=p_{1}E(\left\vert \theta _{1}^{\prime }\right\rangle )+p_{2}E(\left\vert
\theta _{2}^{\prime }\right\rangle ).
\end{equation}%
In matrix notation, $\left\vert \theta _{i}\right\rangle $ can be rewritten
by $\theta _{i}=A_{i}Z\psi _{AB}$. $E(\left\vert \theta _{i}^{\prime
}\right\rangle )$ can be rewritten as
\begin{equation}
E(\left\vert \theta _{i}^{\prime }\right\rangle )=E(U_{i}D_{i}VZY\Lambda
W^{\dag }/\sqrt{p_{i}}),
\end{equation}%
where $\psi _{AB}=Y\Lambda W^{\dag }$ is the singular value decomposition of
$\psi _{AB}$. Since $E$ is invariant under local unitary transformations and
we select $Z=V^{\dag }Y^{\dag }$ for simplicity, eq. (6) can be explicitly
given by
\begin{equation}
E(\left\vert \theta _{1}^{\prime }\right\rangle )=\mathcal{N}\left[ \frac{1}{%
\sqrt{p_{1}}}\left( \sum\limits_{k=1}^{\mathcal{D}}\sqrt{a_{k}\sigma _{k}}%
\right) ^{2}-1\right] .
\end{equation}%
Similarly,
\begin{equation}
E(\left\vert \theta _{2}^{\prime }\right\rangle )=\mathcal{N}\left[ \frac{1}{%
\sqrt{p_{2}}}\left( \sum\limits_{k=1}^{\mathcal{D}}\sqrt{\sqrt{1-a_{k}^{2}}%
\sigma _{k}}\right) ^{2}-1\right] .
\end{equation}%
Substituting eq. (7) and eq. (8) into eq. (5), $\left\langle E(\left\vert
\psi \right\rangle _{AB})\right\rangle $ can be written as%
\begin{eqnarray}
&&\left\langle E(\left\vert \psi \right\rangle _{AB})\right\rangle  \notag \\
&=&\mathcal{N}\left[ \sqrt{p_{1}}\left( \sum\limits_{k=1}^{\mathcal{D}}\sqrt{%
a_{k}\sigma _{k}}\right) ^{2}+\sqrt{p_{2}}\left( \sum\limits_{k=1}^{\mathcal{%
D}}\sqrt{\sqrt{1-a_{k}^{2}}\sigma _{k}}\right) ^{2}-1\right]  \notag \\
&=&\mathcal{N}\left[ \sum\limits_{k=1}^{\mathcal{D}}\left( \sqrt{p_{1}}a_{k}+%
\sqrt{p_{2}}\sqrt{1-a_{k}^{2}}\right) \sigma _{k}\right.  \notag \\
&&+\left. \sum\limits_{i\neq j}\left( \sqrt{p_{1}a_{i}a_{j}}+\sqrt{p_{2}%
\sqrt{\left( 1-a_{i}^{2}\right) \left( 1-a_{j}^{2}\right) }}\right) \sqrt{%
\sigma _{i}\sigma _{j}}-1\right]  \notag \\
&\leq &\mathcal{N}\left[ \left( \sum\limits_{k=1}^{\mathcal{D}}\sqrt{\sigma
_{k}}\right) ^{2}-1\right] =E(\left\vert \psi \right\rangle _{AB}),
\end{eqnarray}%
where the inequality follows from $\sqrt{p_{1}}a_{k}+\sqrt{p_{2}}\sqrt{%
1-a_{k}^{2}}\leq 1$ for any $k$, $\sqrt{p_{1}a_{i}a_{j}}+\sqrt{p_{2}\sqrt{%
\left( 1-a_{i}^{2}\right) \left( 1-a_{j}^{2}\right) }}\leq 1$ and $%
p_{1}^{2}+p_{2}^{2}=1$. Eq. (9) shows that $E$ is an entanglement monotone. $%
\hfill \Box $

\section{\protect\bigskip Genuine tripartite entanglement monotone for
tripartite pure states}

Let us focus on a $\left( 2\otimes 2\otimes n\right) -$dimensional
tripartite quantum pure state $\left\vert \psi \right\rangle _{ABC}$ defined
in the Hilbert space $H_{1}\otimes H_{2}\otimes H_{3}$, the $\left( 2\otimes
2\right) $ reduced density matrix of which can be given by
\begin{equation}
\rho _{AB}=tr_{C}\left[ \left\vert \psi \right\rangle _{ABC}\left\langle
\psi \right\vert \right] .
\end{equation}%
Denote the eigenvalue decomposition of $\rho _{AB}$ by
\begin{equation}
\rho _{AB}=\Phi M\Phi ^{\dag },
\end{equation}%
where the columns of $\Phi $ are the eigenvectors of $\rho _{AB}$ and $M$ is
a diagonal matrix with the diagonal elements being the eigenvalues of $M$.
Define a $4\times 4$ matrix in terms of the spin flip operator $\sigma
_{y}\otimes \sigma _{y}$ as
\begin{equation}
\mathcal{M}=\sqrt{M}\Phi ^{T}\sigma _{y}\otimes \sigma _{y}\Phi \sqrt{M}.
\end{equation}%
Then $\mathcal{M}$ can be regarded as an unnormalized pure state given in
matrix notation analogous to eq. (1). Therefore, the separability of $%
\mathcal{M}$ can be characterized by our bipartite entanglement monotone
introduced in the previous section. Therefore, we have the following theorem.

\textbf{Theorem 2.}-\textit{For a }$\left( 2\otimes 2\otimes n\right) -$%
\textit{dimensional tripartite quantum pure state }$\left\vert \psi
\right\rangle _{ABC}$\textit{, the genuine tripartite entanglement measure
can be given by }%
\begin{eqnarray}
E(\left\vert \psi \right\rangle _{ABC}) &=&E(\mathcal{M})=\mathcal{\tilde{N}}%
\left[ \mathcal{S}\left( \mathcal{MM}^{\dagger }\right) -\mathcal{F}\left(
\mathcal{MM}^{\dagger }\right) \right]  \\
&=&\mathcal{\tilde{N}}\left[ \mathcal{S}\left( \rho _{AB}\tilde{\rho}%
_{AB}\right) -\mathcal{F}\left( \rho _{AB}\tilde{\rho}_{AB}\right) \right]
\\
&=&\mathcal{\tilde{N}}\sum_{i\neq j}\sqrt{\sigma _{i}\sigma _{j}},
\end{eqnarray}%
\textit{where }$\tilde{\rho}_{AB}=$\textit{\ }$\left( \sigma _{y}\otimes
\sigma _{y}\right) \rho _{AB}^{\ast }\left( \sigma _{y}\otimes \sigma
_{y}\right) $\textit{, }%
\begin{equation}
\mathcal{F}\left( \mathcal{MM}^{\dagger }\right) =\sum_{i=1}^{4}\sigma _{i},
\end{equation}%
\textit{with }$\sigma _{i}$\textit{\ being the singular values of }$\mathcal{%
M}$\textit{\ or the eigenvalues of }$\rho _{AB}\tilde{\rho}_{AB}$ \textit{%
and }$\tilde{N}$\textit{\ is an adjustable constant.}

\textbf{Proof. }Note that the equivalence between eq. (13) and eq. (14) is
implied in Ref. [1] and eq. (15) can be easily derived by substituting $%
\sigma _{i}$ into eq. (13) or eq. (14).

First of all, we show that\textbf{\ }$E(\left\vert \psi \right\rangle
_{ABC}) $ characterizes the genuine tripartite entanglement. $\left\vert
\psi \right\rangle _{ABC}$ can also be considered as a $\left( 4\otimes
n\right) $- dimensional bipartite quantum pure state defined in the Hilbert
space $\left( H_{1}\otimes H_{2}\right) \otimes H_{3}$. Based on the Schmidt
decomposition, one can always select a proper basis such that $\left\vert
\psi \right\rangle _{ABC}=\sum\limits_{i=1}^{4}\alpha _{i}\left\vert \tilde{%
\varphi}_{i}\right\rangle \left\vert \check{\varphi}_{i}\right\rangle ,$
where $\left\{ \left\vert \tilde{\varphi}_{i}\right\rangle \right\} $ is the
orthogonal and complete basis of subspace $H_{1}\otimes H_{2}$ and $\left\{
\left\vert \check{\varphi}_{i}\right\rangle \right\} $ is the orthogonal and
complete basis of a $\left( 4\times 4\right) $ -dimensional subspace in $%
H_{3}$. Select some orthogonal and complete basis of $H_{3}$ that must
include $\left\{ \left\vert \check{\varphi}_{i}\right\rangle \right\} $.
Then one can construct an $\left( n\times n\right) $- dimensional matrix $%
\mathcal{\tilde{M}}$ of which $\mathcal{M}$ is a $4\times 4$ block and the
rest is zero. Therefore, $\mathcal{\tilde{M}}$ and $\mathcal{M}$ have the
same entanglement in terms of our bipartite entanglement measure. Most
importantly, the construction of $\mathcal{\tilde{M}}$ is completely
consistent with the '$\mathcal{M}$' introduced in Ref. [12]. That is to say,
the characterization of separability of $\mathcal{M}$ reveals the genuine
tripartite entanglement of $\left\vert \psi \right\rangle _{ABC}$.

Next, we show that $E(\left\vert \psi \right\rangle _{ABC})$ is an
entanglement monotone. We first show that $E(\left\vert \psi \right\rangle
_{ABC})$ does not increase under LOCC in party $A$ only, due to the
invariance of the permutation of party $A$ and $B$. Analogously to Ref. [5]
and the analysis in the previous section, we again consider a sequence of
two-outcome POVM's. Let $\tilde{A}_{1}$ and $\tilde{A}_{2}$ be two POVM
elements such that $\tilde{A}_{1}^{\dag }\tilde{A}_{1}+\tilde{A}_{2}^{\dag }%
\tilde{A}_{2}=\mathbf{1}_{2}$, then $\tilde{A}_{i}=\tilde{U}_{i}\tilde{D}_{i}%
\tilde{V}$, where $\tilde{U}_{i}$ and $\tilde{V}$ are unitary matrices and $%
\tilde{D}_{i}$ are diagonal matrices with entries $(a,b)$ and $[\sqrt{1-a^{2}%
},\sqrt{1-b^{2}}]$, respectively. For an initial tripartite pure state $%
\left\vert \Psi \right\rangle $, let $\left\vert \Theta _{i}\right\rangle
=(A_{i}\otimes \mathbf{1}_{2}\otimes \mathbf{1}_{n})\left\vert \Psi
\right\rangle $ be the unnormalized states obtained after the POVM
operations. The corresponding normalized states can be given by $\left\vert
\Theta _{i}^{\prime }\right\rangle =\left\vert \Theta _{i}\right\rangle /%
\sqrt{p_{i}^{\prime }}$, where $p_{i}^{\prime }=\left\langle \Theta
_{i}\right\vert \left. \Theta _{i}\right\rangle $. Then
\begin{equation}
\left\langle E(\left\vert \Psi \right\rangle )\right\rangle =p_{1}^{\prime
}E(\left\vert \Theta _{1}^{\prime }\right\rangle )+p_{2}^{\prime
}E(\left\vert \Theta _{2}^{\prime }\right\rangle ).
\end{equation}%
Substituting $\left\vert \Theta _{i}^{\prime }\right\rangle $ into eq. (14),
one quickly obtains%
\begin{eqnarray}
\left\langle E(\left\vert \Psi \right\rangle )\right\rangle
&=&abE(\left\vert \Psi \right\rangle )+\sqrt{\left( 1-a^{2}\right) \left(
1-b^{2}\right) }E(\left\vert \Psi \right\rangle )  \notag \\
&\leq &E(\left\vert \Psi \right\rangle ).
\end{eqnarray}

Now, we analogously let $\hat{A}_{1}$ and $\hat{A}_{2}$ be two POVM elements
performed on subsystem $C$ such that $\hat{A}_{1}^{\dag }\hat{A}_{1}+\hat{A}%
_{2}^{\dag }\hat{A}_{2}=\mathbf{1}_{C}$, and $\hat{A}_{i}=\hat{U}_{i}\hat{D}%
_{i}\hat{V}$, where $\hat{U}_{i}$ and $\hat{V}$ are unitary matrices and $%
\hat{D}_{i}$ are diagonal matrices with entries $(\tilde{a}_{1},\tilde{a}%
_{2},\cdots )$ and $[\sqrt{1-\tilde{a}_{1}^{2}},\sqrt{1-\tilde{a}_{2}^{2}}%
,\cdots ]$, respectively. At the same time, we also suppose that $\hat{Z}%
_{i} $ is a local unitary transformation performed on subsystem $C$ after
the operation of $\hat{A}_{i}$. For some initial state $\left\vert \hat{\Psi}%
\right\rangle $, let $\left\vert \hat{\Theta}_{i}\right\rangle =(\mathbf{1}%
_{A}\otimes \mathbf{1}_{B}\otimes \hat{Z}_{i}\hat{A}_{i})\left\vert \hat{\Psi%
}\right\rangle $ be the unnormalized states obtained after the POVM
operations. The corresponding normalized states can be given by $\left\vert
\hat{\Theta}_{i}^{\prime }\right\rangle =\left\vert \hat{\Theta}%
_{i}\right\rangle /\sqrt{p_{i}^{\prime \prime }}$, where $p_{i}^{\prime
\prime }=\left\langle \hat{\Theta}_{i}\right\vert \left. \hat{\Theta}%
_{i}\right\rangle $. Then

\begin{equation}
\left\langle E(\left\vert \hat{\Psi}\right\rangle )\right\rangle
=p_{1}^{\prime \prime }E(\left\vert \hat{\Theta}_{1}^{\prime }\right\rangle
)+p_{2}^{\prime \prime }E(\left\vert \hat{\Theta}_{2}^{\prime }\right\rangle
).
\end{equation}%
It has been proved in Ref. [12] that any local operation $Q$ performed on
party $C$ of $\left\vert \hat{\Psi}\right\rangle $ can be equivalently
described using the $\left( n\times n\right) $- dimensional symmetric $%
\mathcal{\tilde{M}}$ of $\left\vert \hat{\Psi}\right\rangle $ (the nonzero
elements are only limited in a $4\times 4$ block $\mathcal{M}$) given by
\begin{equation}
\mathcal{\hat{M}=}Q^{T}\mathcal{\tilde{M}}^{T}Q.
\end{equation}%
Therefore, after these local operations $E(\left\vert \hat{\Theta}%
_{i}^{\prime }\right\rangle )$ is given by
\begin{equation}
E(\left\vert \hat{\Theta}_{i}^{\prime }\right\rangle )=E\left[ \frac{\hat{V}%
^{T}\hat{D}_{i}^{T}\hat{U}_{i}^{T}\hat{Z}_{i}^{T}\hat{Y}\hat{\Lambda}\hat{Y}%
^{T}\hat{Z}_{i}\hat{U}_{i}\hat{D}_{i}\hat{V}}{p_{i}^{\prime \prime }}\right]
,
\end{equation}%
where $\mathcal{\tilde{M}}=\hat{Y}\hat{\Lambda}\hat{Y}^{T}$ is the singular
value decomposition of $\mathcal{\tilde{M}}$. For simplicity, select $%
Z_{i}=Y^{\ast }U_{i}^{\dag }$, then eq. (21) can be explicitly given by
\begin{equation}
E(\left\vert \hat{\Theta}_{1}^{\prime }\right\rangle )=\mathcal{\tilde{N}}%
\frac{1}{p_{1}^{\prime \prime }}\left[ \left( \sum\limits_{k=1}^{n}\tilde{a}%
_{k}\sqrt{\sigma _{k}}\right) ^{2}-\sum\limits_{k=1}^{n}\tilde{a}%
_{k}^{2}\sigma _{k}\right] ,
\end{equation}%
where $\sigma _{k}$ are the singular values of $\mathcal{\tilde{M}}$.
Similarly,
\begin{equation}
E(\left\vert \hat{\Theta}_{2}^{\prime }\right\rangle )=\mathcal{\tilde{N}}%
\frac{1}{p_{2}^{\prime \prime }}\left[ \left( \sum\limits_{k=1}^{n}\sqrt{1-%
\tilde{a}_{k}^{2}}\sqrt{\sigma _{k}}\right) ^{2}-\sum\limits_{k=1}^{n}\left(
1-\tilde{a}_{k}^{2}\right) \sigma _{k}\right] .
\end{equation}%
Substituting eq. (22) and eq. (23) into eq. (19), $\left\langle E(\left\vert
\hat{\Psi}\right\rangle )\right\rangle $ can be written as%
\begin{align}
& \left\langle E(\left\vert \hat{\Psi}\right\rangle )\right\rangle =\mathcal{%
\tilde{N}}\left[ \sum\limits_{i\neq j}\left( a_{i}a_{j}+\sqrt{\left(
1-a_{i}^{2}\right) \left( 1-a_{j}^{2}\right) }\right) \sqrt{\sigma
_{i}\sigma _{j}}\right]  \notag \\
& \leq \mathcal{\tilde{N}}\sum\limits_{i\neq j}\sqrt{\sigma _{i}\sigma _{j}}%
=E(\left\vert \hat{\Psi}\right\rangle ),
\end{align}%
where $a_{i}a_{j}+\sqrt{\left( 1-a_{i}^{2}\right) \left( 1-a_{j}^{2}\right) }%
\leq 1$ is employed. Eq. (9) and eq. (24) show that $E$ is an entanglement
monotone, hence $E$ is a good entanglement measure for genuine tripartite
entanglement.$\hfill \Box $

Note that eq. (13) is a variational version of eq. (4) for unnormalized pure
states. One may think that $\mathcal{M}$ should be normalized. However,
because $\mathcal{M}$ is only a middle state of the normalized tripartite
pure state $\left\vert \psi \right\rangle _{ABC}$, it can not be normalized
for the same reason given in Ref. [12]. In fact, according to the onion-like
classification of $\left( 2\otimes 2\otimes n\right) $- dimensional quantum
pure states introduced in Refs. [6,8], all the entanglement of the outer
entanglement class can be irreversibly converted to the entanglement of the
inner class. Since the GHZ-type entanglement with local rank$^{[0]}$%
\footnotetext[0]{%
The local rank can be defined as the rank of the reduced density matrix
traced out for all except one party [6].} $(2,2,2)$ is the innermost
tripartite entanglement class, one can consider the GHZ-type inseparability
of $(2,2,2)$ local rank as a minimal element of high-dimensional quantum
entanglement. $E$ measures the genuine tripartite entanglement by collecting
all the minimal elements of GHZ-type inseparability. Thus one can set $%
\mathcal{\tilde{N}}=1$ in the reference frame of $E(\left\vert \Psi
_{GHZ}\right\rangle )=1$, or $\mathcal{\tilde{N}}=1/3$ in the frame of $%
E(\left\vert \Psi _{\max }\right\rangle )=1$. Here
\begin{equation}
\left\vert \Psi _{GHZ}\right\rangle =\frac{1}{\sqrt{2}}\left( \left\vert
000\right\rangle +\left\vert 111\right\rangle \right)
\end{equation}%
is the GHZ state with local rank $(2,2,2)$ and
\begin{equation}
\left\vert \Psi _{\max }\right\rangle =\frac{1}{2}\left( \left\vert
000\right\rangle +\left\vert 011\right\rangle +\left\vert 102\right\rangle
+\left\vert 113\right\rangle \right)
\end{equation}%
is the maximal tripartite entangled state with local rank $(2,2,4)$ of the
outermost class [8].

\begin{figure}[tbp]
\includegraphics[width=8cm]{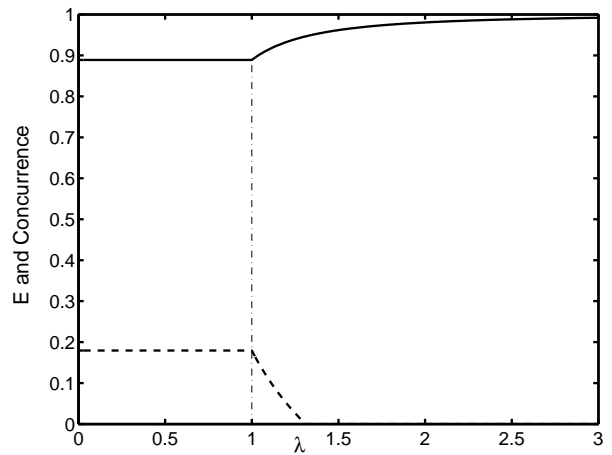}
\caption{Concurrence (dashed line) and genuine tripartite entanglement $E$
(solid line) of the ground state of the isotropic spin-$\frac{1}{2} XY$
chain with three-spin interaction vs $\protect\lambda$. The second-order
quantum phase transition is signaled at $\protect\lambda=1$ by both
entanglement measures, but the singularity at $\protect\lambda=2/(\protect%
\sqrt{2}-1)\protect\pi$ shown by the concurrence does not correspond
to a quantum phase transition. }
\end{figure}

As an application, we consider the connection between the genuine tripartite
entanglement of the ground state and the quantum phase transition of the
isotropic spin-$\frac{1}{2}XY$ chain with three-spin interaction presented
in Refs. [17, 18], which is an exactly solvable quantum spin model. The
Hamiltonian is
\begin{eqnarray}
H &=&-\sum_{i=1}^{N}\left[ \sigma _{i}^{x}\sigma _{i+1}^{x}+\sigma
_{i}^{y}\sigma _{i+1}^{y}\right.  \notag \\
&&\left. +\frac{\lambda }{2}\left( \sigma _{i-1}^{x}\sigma _{i}^{z}\sigma
_{i+1}^{y}-\sigma _{i-1}^{y}\sigma _{i}^{z}\sigma _{i+1}^{x}\right) \right] ,
\end{eqnarray}%
where $N$ is the number of sites, $\sigma _{i}^{\alpha }$($\alpha =x,y,z$)
are the Pauli matrices, and $\lambda $ is a dimensionless parameter
characterizing the three-spin interaction strength. Here the periodic
boundary condition $\sigma _{N+1}=\sigma _{1}$ is assumed. The ground state
of the spin-$\frac{1}{2}XY$ chain can always be considered as a tripartite $%
\left( 2\otimes 2\otimes \left[ 2N-4\right] \right) -$dimensional pure state
by a grouping such as \textit{two-nearest-neighbor-particle} vs. \textit{%
others}.\textit{\ }One can safely employ $E$ to measure the genuine
tripartite entanglement. The \textit{two-nearest-neighbor-particle }density
matrix can be given [18] by
\begin{equation}
\rho _{i,i+1}=\left(
\begin{array}{cccc}
\frac{(1-G^{2})}{4} & 0 & 0 & 0 \\
0 & \frac{(1+G^{2})}{4} & \frac{G}{2} & 0 \\
0 & \frac{G}{2} & \frac{(1+G^{2})}{4} & 0 \\
0 & 0 & 0 & \frac{(1-G^{2})}{4}%
\end{array}%
\right)
\end{equation}%
in the standard basis \{$\left\vert \uparrow \uparrow \right\rangle
,\left\vert \uparrow \downarrow \right\rangle ,\left\vert \downarrow
\uparrow \right\rangle ,\left\vert \downarrow \downarrow \right\rangle $\},
where
\begin{equation}
G=\left\{
\begin{array}{cc}
\frac{2}{\pi }, & \lambda <1, \\
\frac{2}{\pi \lambda }, & \lambda \geq 1.%
\end{array}%
\right.
\end{equation}%
Using eq. (4), one can easily calculate the genuine tripartite entanglement
shown in Fig. 1. It is obvious that the first derivative of $E$ is
discontinuous at $\lambda =1$ which consistent with Ref. [17] shows that the
three-spin interaction leads to a second-order quantum phase transition. $E$
and its first derivative do not show any other singularity, which implies
that $E$ faithfully signals a quantum phase transition. But the first
derivative of the ground-state concurrence of two nearest-neighbor spins
yields another discontinuity at $\lambda =2/(\sqrt{2}-1)\pi $ shown in Fig.
1, implying that the concurrence is misleading for this model.

\section{Discussion and conclusion}

We have introduced an entanglement monotone to measure the genuine
tripartite entanglement existing in a given tripartite $(2\otimes 2\otimes
n)-$dimensional quantum pure states in terms of a new bipartite entanglement
measure. It is a new method to characterize genuine tripartite entanglement
because it collects the contribution of all GHZ-type entanglement.
Furthermore, it is interesting that the squared genuine tripartite
entanglement monotone is the same as the original residual entanglement [9]
for $(2\otimes 2\otimes 2)-$dimensional systems. The extension to mixed
states (including the bipartite entanglement monotone) is straightforward in
principle based on the convex roof construction [19], but an operational
lower bound seems to be a bit difficult which is left to our forthcoming
works. As an application, we considered the genuine tripartite entanglement
of the ground state of the exactly solvable isotropic spin-$\frac{1}{2}$
chain with three-spin interaction. It is shown that the singularity of the
genuine tripartite entanglement exactly signals a quantum phase transition.
However, we only considered the given grouping method. The other $2\otimes
2\otimes 2\left[ N-2\right] $ groupings can also be considered. Then, the
tripartite entanglement monotone may be further employed to analytically
study the quantum phase transition of more physical systems.

\section{Acknowledgement}

This work was supported by the National Natural Science Foundation
of China, under Grant No. 10747112, No. 10575017 and No. 10774020.

\end{document}